\documentclass[useAMS,usenatbib]{mn2e}

\usepackage[dvips]{graphicx}
\usepackage[cmex10]{amsmath}
\usepackage{array}
\usepackage{mdwmath}
\usepackage{mdwtab}

\title[Detection/estimation of polarization data]{Detection/estimation
   of the modulus of a vector. Application to point source detection
   in polarization data}

\author[Arg\"ueso et al.]{F.~Arg\"ueso$^{1}$\thanks{E-mail:
    argueso@uniovi.es}, J.~L.~Sanz$^{2,3}$, D.~Herranz$^{2}$,
  M.~L\'opez-Caniego$^{4}$, and J.~Gonz\'alez-Nuevo$^{5}$ \\ $^{1}$
  Departamento de Matem\'aticas, Universidad de Oviedo, 33007, Oviedo,
  Spain \\ $^{2}$ Instituto de F\'\i sica de Cantabria, CSIC-UC,
  Av. los Castros s/n, Santander, 39005, Spain \\ $^{3}$ CNR Istituto
  di Scienza e Tecnologie dell'Informazione, via G. Moruzzi 1,
  I-56124, Pisa, Italy \\ $^{4}$ Astrophysics Group, Cavendish
  Laboratory, J.J. Thomson Avenue, CB3 0HE, Cambridge, United Kingdom
   \\ $^{5}$ SISSA-I.S.A.S., via
  Beirut 4, I-34014 Trieste, Italy}

\begin{document}

\date{Received --, Accepted --}

\pagerange{\pageref{firstpage}--\pageref{lastpage}} \pubyear{2008}

\maketitle

\label{firstpage}

\maketitle

\begin{abstract}
Given a set of images, whose pixel values can be considered as the
components of a vector, it is interesting to estimate the modulus of
such a vector in some localised areas corresponding to a compact
signal. For instance, the detection/estimation of a polarized signal
in compact sources immersed in a background is relevant in some
fields like astrophysics. We develop two different techniques, one
based on the Neyman-Pearson lemma, the Neyman-Pearson filter (NPF),
and another based on prefiltering-before-fusion, the filtered fusion
(FF), to deal with the problem of detection of the source and
estimation of the polarization given two or three images
corresponding to the different components of polarization (two for
linear polarization, three including circular polarization). For the
case of linear polarization, we have performed numerical simulations
on two-dimensional patches to test these filters following two
different approaches (a blind and a non-blind detection),
considering extragalactic point sources immersed in cosmic microwave
background (CMB) and non-stationary noise with the conditions of the
70 GHz \emph{Planck} channel. The FF outperforms the NPF, especially
for low fluxes. We can detect with the FF extragalactic sources in a
high noise zone with fluxes $\geq (0.42,0.36)$ Jy for
(blind/non-blind) detection and in a low noise zone with fluxes
$\geq (0.22,0.18)$ Jy for (blind/non-blind) detection with low
errors in the estimated flux and position.
\end{abstract}


\section{Introduction} \label{sec:intro}

The detection and estimation of the intensity of compact objects (or
small regions) embedded in a background plus instrumental noise is
relevant in different contexts, e.g. astrophysics, cosmology,
medicine, teledetection, radar, etc. Different techniques have
proven useful in the literature.  Some of the proposed techniques
are frequentist, such as the standard matched filter
\citep{MF_radio92,MF_ROSAT95,MF_CfA97,tegmark98,sanz01,herr02c,MF_poisson06},
the matched multifilter~\citep[or multifrequency
filter,][]{herr02a,herr05} or the recently developed matched matrix
filters~\citep{herranz08a,herranz08b} that correspond to scalar,
vector or matrix filters, respectively. Other frequentist techniques
include continuous wavelets like the standard Mexican
Hat~\citep{vielva01,vielva03,MHW_SCUBA,wsphere} and other members of
its family~\citep{MHW206} and, more generally, filters based on the
Neyman-Pearson approach using the distribution of
maxima~\citep{can05a,can05b}. All of these filters have been used in
the literature, in particular, for the detection and estimation of
the intensity of point-like sources (i.e. extragalactic objects) in
Cosmic Microwave Background (CMB) maps. In addition, some of them
have been applied to real data like those obtained by the WMAP
satellite~\citep{NEWPS07} and simulated
data~\citep{can06,challenge08} for the upcoming experiment on board
the \emph{Planck} satellite~\citep{planck_tauber05}. Besides,
Bayesian methods have also been recently
developed~\citep{hob03,psnakesI}.

In some applications it is important to measure not only the
intensity of the light (signal) but also its polarization. An
example is the study of CMB radiation. The polarization is given by
the Stokes parameters $Q, U, V$, and the total intensity of
polarization is $P \equiv (Q^2+U^2+V^2)^{1/2}$~\citep{kamion97}. For
linear polarization, the previous expression reduces to $P \equiv
(Q^2+U^2)^{1/2}$. In such cases, three or two images are added
quadratically followed by a square root.

Let us consider the case of linear polarization. In this case we
have two images $Q, U$ and different approaches can be used to deal
with detection/estimation of point-like sources on these maps. On
the one hand, one can try to get the polarization $P$ directly on
the $P$-map. In this approach, we will consider one filter, obtained
through the Neyman-Pearson technique (NPF). On the other hand, we
can operate with two matched filters, each one on $Q$ and $U$
followed by a quadratic fusion and square root. We will call this
procedure filtered fusion (FF). It is clear that from a formal point
of view we are trying to ask which is the optimal way to find the
modulus of a vector given the components. In the case we have only
the map of the modulus of a vector and the components are unknown,
the FF cannot be applied and the only possibility is the NPF.

We will develop the methodology for the cases of a 2-vector and a
3-vector because of the possible interesting applications to the
2-plane and 3-space . We will show the results when using numerical
simulations on flat patches that are relevant for the component
separation of sources (linear-polarized extragalactic sources) in
CMB maps.

In section~\ref{sec:methodology}, we develop the methodology. In
section~\ref{sec:simulations}, we describe the numerical simulations
done in order to test the previous techniques. In
section~\ref{sec:results} we present the main results and in
section~\ref{sec:conclusions} we give the main conclusions.

\section{Methodology} \label{sec:methodology}

\subsection{2-vector} \label{sec:method_2vec}

To develop our methodology, we shall assume that we have a compact
source, located for simplicity at the centre of two images $Q$, $U$
and characterised by amplitudes $A_Q$, $A_U$ and a profile $\tau
(\vec{x})$, immersed in noises $n_{Q}(\vec{x})$ and $n_{U}(\vec{x})$
that are Gaussian and independently distributed with zero mean and
dispersions $\sigma_Q(\vec{x})$ and $\sigma_U(\vec{x})$. Again, for
simplicity, we will consider that
$\sigma_Q(\vec{x})=\sigma_U(\vec{x})=\sigma(\vec{x})$, a condition
that is verified in most polarization detectors. In general, we will
consider that the noise is non-stationary. We remark that the
previous assumptions can be easily generalised to different profiles
and different types of noise in the two images but we will not
consider it in this paper. We will assume a linear model for the two
images

\begin{equation} \label{eq:datamodel}
  d_{Q,U}(\vec{x}) = A_{Q,U}\tau (\vec{x}) + n_{Q,U}(\vec{x}).
\end{equation}

The $P$-map, $P(\vec{x})\equiv (Q^2(\vec{x})+U^2(\vec{x}))^{1/2}$,
is characterised by a source at the centre of the image with
amplitude $A\equiv (A_Q^2+A_U^2)^{1/2}$ immersed in non-additive
noise which is correlated with the signal.

\subsubsection{Neyman-Pearson filter (NPF) on the $P$-map}

If the noise is distributed normally and independently, then at any
point the integration over the polar angle leads to the 2D-Rayleigh
distribution of $P$ in absence of the source~\citep{papoulis}
\begin{equation} \label{eq:rayleigh-p}
  f(P|0) = \frac{P}{\sigma^2}e^{-P^2/2\sigma^2},
\end{equation}
\noindent whereas if the source is present, with amplitude $A$, one
obtains the Rice distribution~\citep{Rice}
\begin{equation} \label{eq:rice-p}
  f(P|A) = \frac{P}{\sigma^2}e^{-(A^2 + P^2)/2\sigma^2}
  I_o\left(A\frac{P}{\sigma^2}\right),
\end{equation}
\noindent where $I_o$ is the modified Bessel function of zero order.
If our image is pixelised, the different data $P_i$, $i=1,\ldots,
n$, with $n$ the number of pixels, will follow the two distributions
\begin{eqnarray}
  (H_0): \ & f(P_i|0) =
  \prod_i\frac{P_i}{\sigma_i^2}e^{-\sum_iP_i^2/2\sigma_i^2} \\ (H_1):
  \ & f(P_i|A) =
  \prod_i\frac{P_i}{\sigma_i^2}I_o(A\frac{P_i\tau_i}{\sigma_i^2})
  e^{-\sum_i(A^2\tau_i^2 + P_i^2)/2\sigma_i^2},
\end{eqnarray}
\noindent being $\sigma_i$ the value of $\sigma$ in the
$i^{\mathrm{th}}$ pixel, $H_0$, $H_1$ the null hypothesis (absence of
source) and the alternative (presence of source), respectively, and
$\tau_i$ the profile at the $i^{\mathrm{th}}$ pixel. The
log-likelihood is defined by
\begin{eqnarray}
  l(A|P_i) = \log \frac{f(H_1)}{f(H_0)} =
  \nonumber\\ -A^2\sum_i\frac{\tau_i^2}{2\sigma_i^2}+ \sum_i
  \log\,I_o\left(A\frac{P_i\tau_i}{\sigma_i^2}\right). \label{eq:log_lik2}
\end{eqnarray}
\noindent The maximum likelihood estimator of the amplitude,
$\hat{A}$, is given by the solution of the equation
\begin{equation} \label{eq:mle}
  \hat{A} \sum_i\frac{\tau_i^2}{\sigma_i^2} =
  \sum_iy_i\frac{I_1(\hat{A}y_i)}{I_o(\hat{A}y_i)},\ \ \ y_i\equiv
  \frac{P_i\tau_i}{\sigma_i^2}.
\end{equation}
\noindent This equation can be interpreted as a non-linear filter
operating on the data that we shall call the Neyman-Pearson filter
(NPF).

\subsubsection{Filtered fusion (FF)}

In this case we use the same matched filter (MF) operating over the
two images $Q$, $U$, respectively, as given by~\citep{paco08}:

\begin{equation} \label{eq:MF_nonstat}
  \Phi (\vec{x})\propto \frac{\tau (\vec{x})}{\sigma^2(\vec{x})}.
\end{equation}

 \noindent Then, with the two filtered images $Q_{MF}, U_{MF }$ we make the
non-linear fusion $P \equiv (Q_{MF}^2+U_{MF}^2)^{1/2}$ pixel by
pixel.

\subsection{3-vector} \label{sec:method_3vec}

Now, we shall assume that we have a compact source at the centre of
three images $Q$, $U$, $V$ characterised by amplitudes $A_Q$, $A_U$,
$A_V$ and a profile $\tau (\vec{x})$ immersed in noise $n_{Q,U,V}
(\vec{x})$ that is Gaussian and independently distributed with zero
mean and dispersion $\sigma(\vec{x})$. In general, we will consider
that the noise is non-stationary. We will assume a linear model for
the three images
\begin{equation} \label{eq:model3}
  d_{Q,U,V}(\vec{x}) = A_{Q,U,V}\tau (\vec{x}) + n_{Q,U,V}(\vec{x}).
\end{equation}
The $P$-map, $P(\vec{x})\equiv
(Q^2(\vec{x})+U^2(\vec{x})+V^2(\vec{x}))^{1/2}$, is characterised by
a source at the centre of the image with amplitude $A\equiv
(A_Q^2+A_U^2+A_V^2)^{1/2}$ immersed in non-additive noise correlated
with the signal.

\subsubsection{Neyman-Pearson filter (NPF) on the $P$-map}

If the noise is distributed normally and independently, then at any
point the integration over the angles leads to the 3D-Rayleigh
distribution in absence of the source, also called the
Maxwell-Boltzmann distribution in Physics
\begin{equation}   \label{eq:rayleigh3}
  f(P|0) =
  \left(\frac{2}{\pi}\right)^{1/2}\frac{P^2}{\sigma^3}e^{-P^2/2\sigma^2},
\end{equation}
\noindent whereas if the source is present, with amplitude $A$, one
obtains the distribution
\begin{equation} \label{eq:rice3}
  f(P|A) = \left(\frac{2}{\pi}\right)^{1/2}\frac{P}{\sigma A}e^{-(A^2
    + P^2)/2\sigma^2} \sinh\left(\frac{AP}{\sigma^2}\right).
\end{equation}
\noindent If our image is pixelised, the different data $P_i$, $i=
1,\ldots, n$ follow the two distributions
\begin{eqnarray}
(H_0): & f(P_i|0) = \left(\frac{2}{\pi}\right)^{n/2} \prod_i
  \frac{P_i^2}{\sigma_i^3}e^{- \sum_iP_i^2/2\sigma_i^2} \\ (H_1): &
  f(P_i|A) = \left(\frac{2}{\pi}\right)^{n/2} \prod_i
  \frac{P_i}{\sigma_i A\tau_i} \times \nonumber \\ &
  \sinh\left(A\frac{P_i\tau_i}{\sigma_i^2}\right)
  e^{-\sum_i(A^2\tau_i^2 + P_i^2)/2\sigma_i^2},
\end{eqnarray}
\noindent being $H_0$, $H_1$ the null hypothesis (absence of source)
and the alternative (presence of source), respectively and $\tau_i$
the profile at the $i^{\mathrm{th}}$ pixel. The log-likelihood is
defined by
\begin{eqnarray}
  l(A|P_i) = \log \frac{f(H_1)}{f(H_0)} =
  -A^2\sum_i\frac{\tau_i^2}{2\sigma_i^2} \nonumber \\ - n\log A+
  \sum_i\log \left[\sinh
    \left(A\frac{P_i\tau_i}{\sigma_i^2}\right)\right].
  \label{eq:log_likelihood3}
\end{eqnarray}
\noindent The maximum likelihood estimator of the amplitude,
$\hat{A}$, is given by the solution of the equation
\begin{equation}
\hat{A}\sum_i\frac{\tau_i^2}{\sigma_i^2} + \frac{n}{\hat{A}}=
\sum_iy_i\coth(\hat{A}y_i),\ \ \ y_i\equiv
\frac{P_i\tau_i}{\sigma_i^2}.
\end{equation}
\noindent This equation can be interpreted as a non-linear filter
operating on the data that we shall also call the Neyman-Pearson
filter (NPF).

\subsubsection{Filtered fusion (FF)}

In this case we use the same MF operating over the three images $Q$,
$U$, $V$, respectively, as given by equation (\ref{eq:MF_nonstat}).
Then, with the three filtered images $Q_{MF}$, $U_{MF }$, $V_{MF }$
we make the non-linear fusion $P\equiv
(Q_{MF}^2+U_{MF}^2+V_{MF}^2)^{1/2}$ pixel by pixel.

\section{Simulations} \label{sec:simulations}

The European Space Agency (ESA) \emph{Planck}
satellite~\citep{planck_tauber05}, to be launched in 2009, will
measure the anisotropies of the CMB with unprecedented accuracy and
angular resolution. It will also analyse the polarization of this
primordial light. It is of great interest the detection and
estimation of polarized sources in CMB
maps~\citep{bluebook,tucci04,tucci05}; since this radiation is
linearly polarized~\citep{kamion97}, $V=0$, we will apply the
methods for the detection/estimation of the modulus of a 2-vector
presented in section~\ref{sec:method_2vec}. However, some
cosmological models predict a possible circular polarization of CMB
radiation~\citep[see for example][]{cooray03,agar08}. Even if CMB is
not circularly polarized, the extragalactic radio sources can indeed
show circular polarization~\citep{aller05,kirk06,homan06}.  Besides,
circular polarization occurs in many other astrophysical areas, from
Solar Physics~\citep{tris07,reiner07} to interestellar
medium~\citep{cox07}, just to put a few examples.  In all these
cases the results for the modulus of a 3-vector presented in
section~\ref{sec:method_3vec} could be useful.

In order to compare and evaluate the performance of the filters
presented in section~\ref{sec:method_2vec}, we have carried out
two-dimensional simulations with the characteristics of the 70 GHz
\emph{Planck} channel~\citep{bluebook}. The simulated images have
$16\times16$ pixels with a pixel size of $3.43$ arcmin. We simulate
the $Q$ and $U$ components of the linear polarization as follows:
each component consists of Gaussian uncorrelated detector noise plus
the contribution of the CMB and a polarized point source filtered
with a Gaussian-shaped beam whose full width at half maximum (FWHM)
is $14'$, (the FWHM of the 70 GHz \emph{Planck} channel beam). So
the source polarization components can be written as
\begin{eqnarray}
  s_Q & \equiv & A_Q \exp \left[ -\frac{|\vec{x}|^2}{2\gamma^2}
    \right] \\ s_U & \equiv & A_U \exp \left[
    -\frac{|\vec{x}|^2}{2\gamma^2} \right],
\end{eqnarray}
\noindent where $\gamma$ is the beam dispersion (size) and we assume
that the source is centred at the origin.  The CMB simulation is based
on the observed WMAP low multipoles, and on a Gaussian realisation
assuming the WMAP best-fit $C_{\ell}$ at high multipoles.  We do not
include other possible foregrounds, since we are doing a first
approximation to the detection/estimation problem and we assume that
we apply our filters to relatively clean areas far away enough from
the Galactic plane. Alternatively, we could consider a case in which
the majority of foregrounds have been previously removed by means of
some component separation method.

We consider realistic non-stationary detector noise.  We have
simulated the noise sky pattern for a \emph{Planck} flight duration of
14 months, assuming a simple cycloidal scanning strategy with a 7
degree slow variation in the Ecliptic collatitude of the
$z$-axis. This scanning strategy implies that the sky will be covered
inhomogeneously. The simulations have the same characteristics as the
ones used in~\citet{paco06}. In order to illustrate the effects of
non-stationary noise we have chosen two representative zones of the
sky: one zone of high noise but quite isotropic and another zone of
low noise but more anisotropic. The first zone is located in a region
far from the Ecliptic poles, where the noise pattern is quite uniform
and the number of hits per pixel of the detector is small.  The
average r.m.s deviation of the first zone (high noise) in units of
$\Delta T/T $ (thermodynamic) is $\sigma=3\times 10^{-5}$ (for each
component $Q$ and $U$) and its standard deviation is $4.2\times
10^{-6}$.  For this particular scanning strategy approximately $25\% $
of the sky has this kind of noise pattern.  The second zone is close
to one of the Ecliptic poles, where the sky is scanned more times (low
noise level) but the hit pattern is very inhomogeneous.  It has an
average r.m.s.  deviation $\sigma=1.1\times 10^{-5}$ and its standard
deviation is $3.8\times 10^{-6}$. Then, in the second zone the noise
is lower but proportionally more anisotropic. For this scanning
strategy, $\sim 6\%$ of the sky has a noise pattern with these
characteristics.  Other zones of the sky would be intermediate
cases between those considered here.

We take values of the source fluxes (before filtering with the
Gaussian beam) $F_Q$ and $F_U$ ranging from $0.1$ Jy to $0.5$ Jy
with a step of $0.1$ Jy. A flux of 0.1 Jy corresponds for the 70 GHz
channel to $\Delta T/T=1.9\times 10^{-5}$, so that it is of the
order of the detector noise r.m.s deviation. The fluxes also
correspond to typical polarization fluxes~\citep{tucci04,tucci05}.
As we will see in section~\ref{sec:results}, for some low-noise
cases it is necessary to simulate even lower flux sources in order
to study the behaviour of the filters in the low signal to noise
ratio regime. We will explicitly report on this in the appropriate
cases in section~\ref{sec:results}. The number of simulations is
$500$ for each combination of pairs of values of $F_Q$ and $F_U$.
After carrying out the corresponding simulations for $Q$ and $U$, we
add them quadratically and take the square root to calculate
$P=\sqrt{Q^2+U^2}$, the polarization modulus.

\begin{figure*}
\centering
\includegraphics{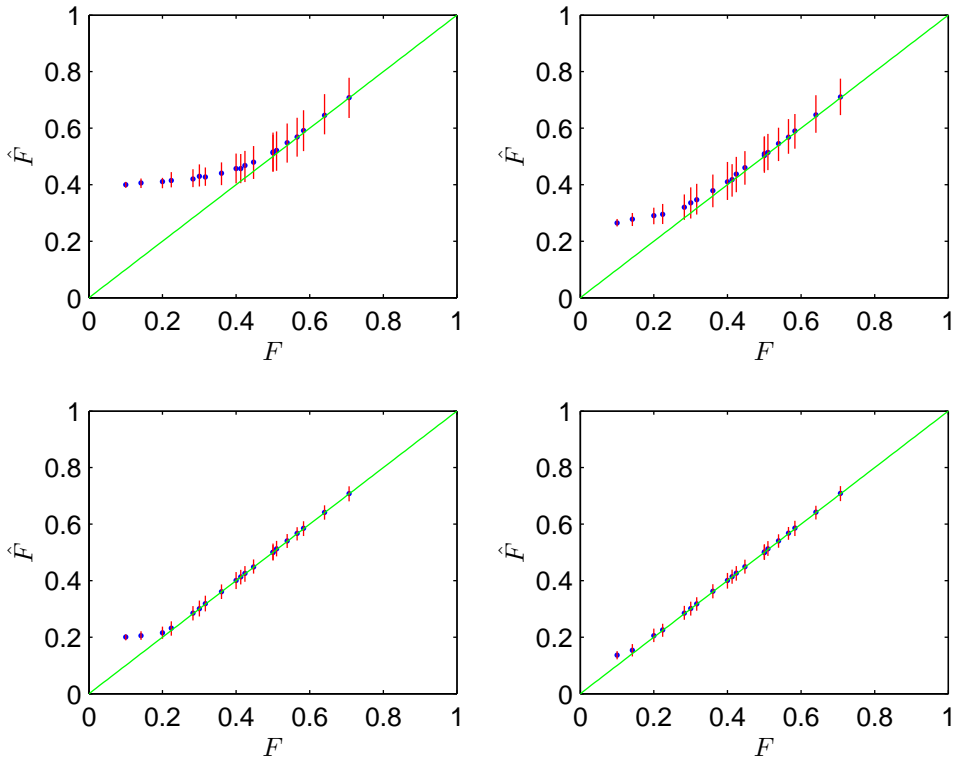}
\caption{Blind detection. Estimated flux $\hat{F}$ in Jy of the
  polarized sources plotted against their real flux $F$. The average
  and $68\%$ confidence intervals (vertical bar) of 500 simulations
  are plotted.  Top left: the NPF has been used and the noise
  corresponds to the high noise zone. Top right: FF and high
  noise. Bottom left: NPF and low noise . Bottom right: FF and low
  noise. In all the plots the straight line $\hat{F}=F$ is drawn for
  comparison.}
\label{fig:figpola1}
\end{figure*}

\begin{figure*}
\centering
\includegraphics{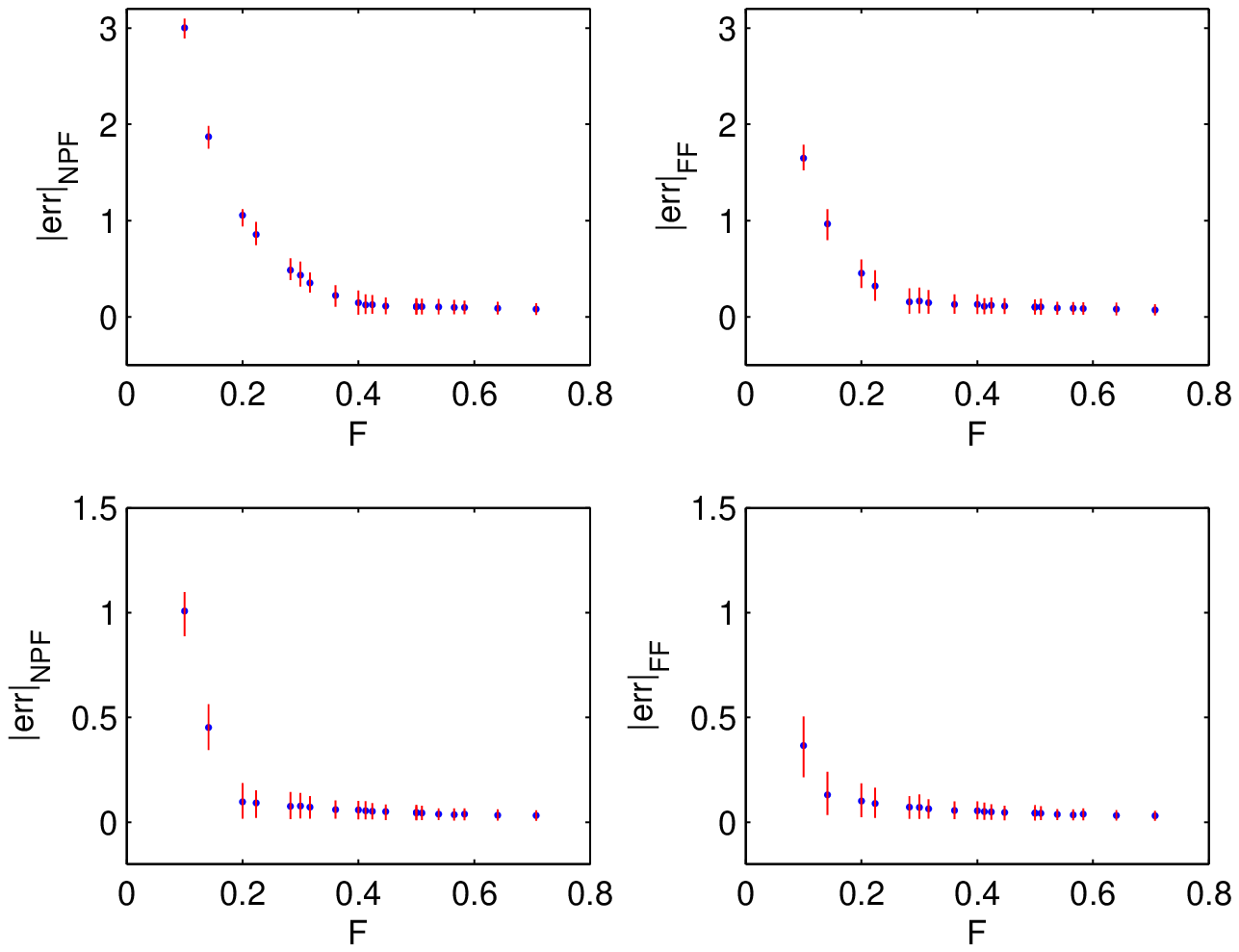}
\caption{Blind detection. Absolute value of the relative error,
   $|$err$|$, in the estimation of the flux of polarized sources
   plotted against the real flux $F$ in Jy. The average and $68\%$
   confidence intervals (vertical bar) of 500 simulations are
   plotted. Top left: The NPF is used with high noise. Top right: FF
   and high noise. Bottom left: NPF and low noise. Bottom right: FF
   and low noise.}
\label{fig:figpola2}
\end{figure*}

We study two different detection types: blind detection and
non-blind detection. In the former case, we assume that we do not
know the position of the source and then we place it at random in
the image, in the latter case we know the source position and then
we place it at the centre of the patch. We have considered images of
$16\times 16$ pixels in order to do fast calculations. In order to
avoid border effects, we simulate and filter $24\times 24$ pixel
patches and, after the filtering step, we retain only the $16\times
16$ pixel central square. We have tested the case of larger patches
but the obtained results do not change significantly.

We use two different filters: the filtered fusion (FF), which consists
in the application of the matched filter to the images in $Q$ and $U$
separately in a first moment, and then the calculation/construction of
$P$ from the matched-filtered images $Q_{MF}$ and $U_{MF}$ and the
Neyman-Pearson filter (NPF), applied directly on $P$, derived from the
Neyman-Pearson lemma and presented with detail in
section~\ref{sec:methodology}. These filters are suitable for the case
of uncorrelated noise, but we apply them to simulations including the
CMB, which is correlated. However, we have checked that our results
are similar when we consider simulations with and without the CMB: the
relative differences of the errors in the estimated fluxes and
positions are at most of a few percent.  This is due to the low value
of the CMB r.m.s. deviation $\sigma \simeq 6\times 10^{-7}$ as
compared to that of the detector noise. Hence, the methods derived in
section~\ref{sec:method_2vec} are also suitable for simulations
including the CMB in the 70 GHz \emph{Planck} case we consider.

In the blind case we apply these filters to each simulation, centering
the filters successively at each pixel, since we do not know the
source position.  We estimate the source amplitude $A$ for the NPF, in
this case we calculate the value of $A$ which maximises the
log-likelihood, eqs. (\ref{eq:log_lik2}) and (\ref{eq:mle}). For the
FF we estimate separately $A_Q$ and $A_U$ and obtain
$A=\sqrt{A_Q^2+A_U^2}$.

We compute the absolute maximum of $A$ in each filtered map and keep
this value as the estimated value of the polarization $P$ of the
source and the position of the maximum as the position of the
source. Note that in the more realistic case where more than one
source can be present in the images, it is still possible to proceed
as described by looking for local peaks in the image. In the
non-blind case we only centre the filters in the pixels included in
one FWHM of the source centre (approximately $10\%$ of the total).
In this way, we use the knowledge of the source position, then we
calculate the absolute maximum of the estimated $A$ in these pixels.
We also calculate the significance level of each detection.  In
order to do this, we carry out 1000 simulations with $A_Q=0$ and
$A_U=0$, and we calculate the estimated value of the source
polarization in this case for each filter. We consider the null
hypothesis, $H_0$, there is no polarized source, against the
alternative hypothesis, $H_1$, there is a polarized source. We set a
significance level $ \alpha =0.05 $, this means that we reject the
null hypothesis when a simulation has an estimated source amplitude
higher than that of $95\%$ of the simulations without polarized
source. We define the power of the test as $1-\delta$, with $\delta$
the probability of accepting the null hypothesis when it is false,
i.e. the power is the proportion of simulations with polarized
source with an estimated amplitude higher than that of $95\%$ of the
simulations without source. The higher the power the more efficient
the filter is for detection.  Note that the test can be performed in
the same way in the blind and non-blind cases: in the second case we
know the position of the source, but we do not know whether it is
polarized or not.

\section{Results} \label{sec:results}

\subsection{Blind case} \label{sec:blind}

We carry out simulations in the blind case for the high noise and
low noise zones as explained above. We apply the filters to the
images and calculate the absolute maximum of $A$ for each filtered
image. If the detection has a significance higher than $\alpha=0.05$
we consider it as a real detection, otherwise it would be a spurious
one. For this significance level, we calculate the power of the
detection test for the different pairs of $F_Q$ and $F_U$ values. We
also calculate the estimated value of the polarization amplitude
$\hat{A}$, convert it to the estimated flux in Jy, $\hat{F}$, and
compare it with the real value $F=\sqrt{F_{Q}^2+F_{U}^2}$. We plot
$\hat{F}$ against $F$ for the NPF and the FF with high and low noise
in Fig.~\ref{fig:figpola1}. We also compute the relative error of
this estimation and its absolute value. For each pair of $F_Q$ and
$F_U$ values, we obtain the average and $68\%$ confidence intervals
of the absolute value of the relative error, taking into account the
$500$ performed simulations. These values are plotted against $F$ in
Fig.~\ref{fig:figpola2} for the same cases shown in
Fig.~\ref{fig:figpola1}. We also calculate the estimated position of
the detected source and obtain the position error expressed in terms
of the number of pixels. The power and the average of the relative
error in $F$, of its absolute value and of the position error are
presented in Table~\ref{tb:table1} for the two different types of
noise. The rows in the Table are sorted in ascending order of $F$.
For the high noise case, we see in the Table that the power for the
FF is higher than for the NPF. The position and polarization errors
are also lower for the FF, this can also be seen in
Figs.~\ref{fig:figpola1} and~\ref{fig:figpola2}. For $F >0.6$ Jy,
the FF and NPF perform similarly. For fluxes higher than $0.42$ Jy,
the power with the FF is $ = 100\%$, the average relative error
(bias) is $\leq 0.03$, the average of the absolute value of the
relative error is $\leq 0.12$ and the average position error is
$\leq 0.36$. From now on, we will use the flux limit for which the
power is $100\%$ as a measure of the filter performance.

\begin{figure*}
\centering
\includegraphics{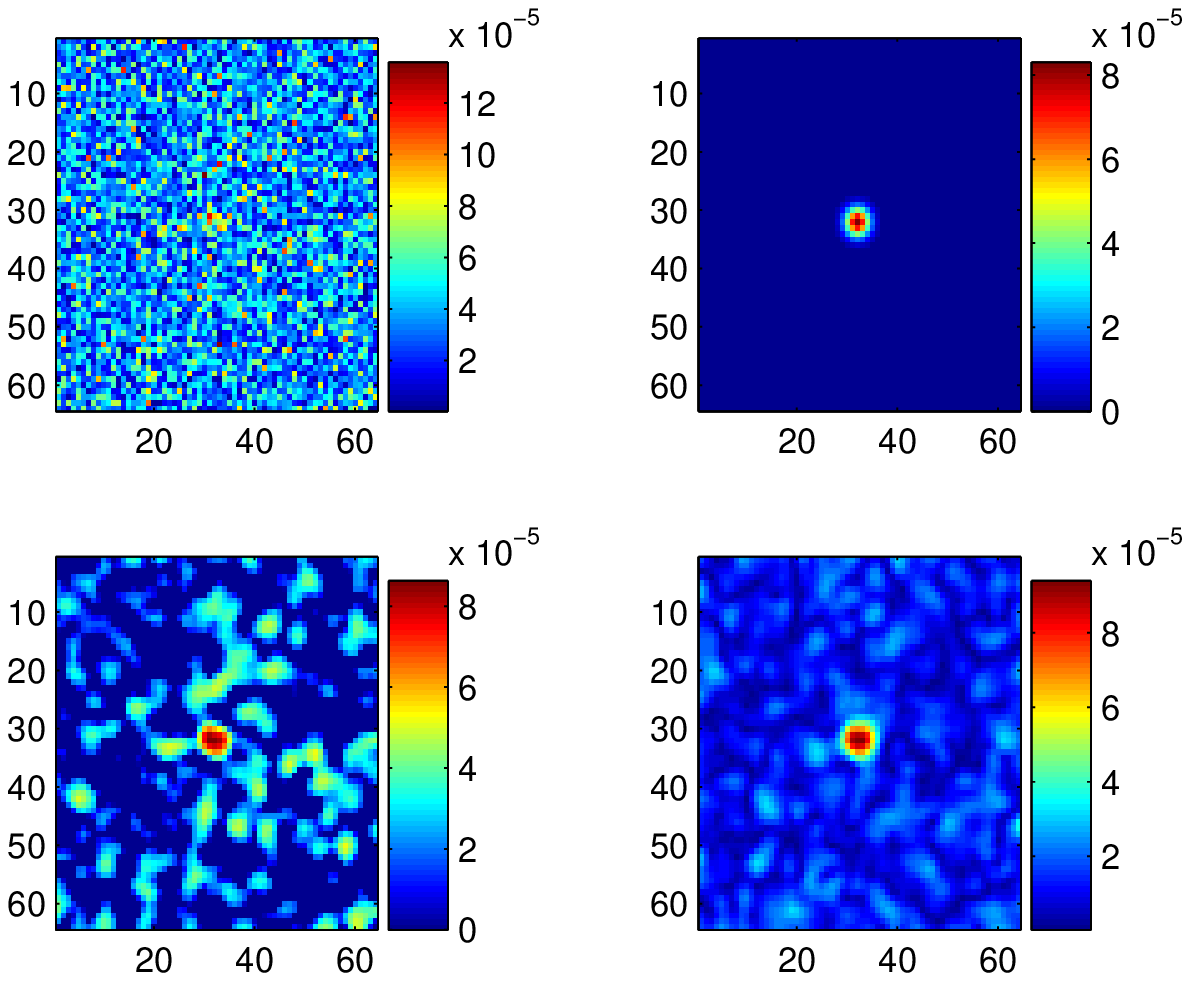}
\caption{Blind detection image. Top left: Image of a polarized
source
  filtered with a Gaussian beam (FWHM=14') placed in the centre of an
  image of $64\times 64$ pixels with pixel size=$3.43'$. The source
  polarization flux components are $(F_Q,F_U)=(0.4,0.4)$ in Jy and it
  is embedded in CMB plus detector noise (high noise zone). Top right:
  The image of the polarized source only. Bottom left: The first image
  after the application of the NPF. Bottom right: The first image
  after the application of the FF.}
\label{fig:figpola3}
\end{figure*}

The results for the low noise case are logically much better, the FF
also performs better than the NPF. The power of the two filters
quickly reaches 100$\%$ for fluxes $F>0.28$ Jy and the errors in
both flux and position remain stable from fluxes $F\geq 0.40$ Jy. We
therefore have cut the Table at $F=0.40$ Jy. In order to have a
better sampling of the interesting signal to noise regime, we have
simulated (using the same number of simulations as in the other
cases) in addition the flux pairs $(F_Q,F_U)=(0.05,0.10)$,
$(0.00,0.15)$, $(0.05,0.15)$, $(0.10,0.15)$ and $(0.00,0.25)$ Jy.
This way the Table is much more informative. For fluxes higher than
$0.22$ Jy, the power with the FF is $100\%$, the average bias is
$\leq 0.01$, the average of the absolute value of the relative error
is $\leq 0.09$ and the average position error is $\leq 0.21$.

The FF is also much faster than the NPF. For instance the analysis of
an image of 64x64 pixels in a personal computer takes 7 seconds with
the FF and about 6 minutes with the NPF. This is due to the
maximization involved. The computation time grows proportionally to
the number of pixels, so that the NPF could be too slow if we want to
analyze large images. This is the reason why we have considered small
patches.

Finally, we show in Fig.~\ref{fig:figpola3} four images
corresponding to a polarized source with $(F_Q,F_U)=(0.4,0.4)$
embedded in high noise. For the sake of a better visualisation, we
show $64\times64$ pixel images instead of the $16\times16$ sized
images used in the simulations. We show the original image in $P$
including noise, CMB and source, the image of the source, the image
filtered with the NPF and the image treated with the FF method. By
simple visual inspection, it is possible to see that the the FF
performs better than the NPF.

\subsection{Non-blind case} \label{sec:nonblind}

\begin{figure*}
\centering
\includegraphics{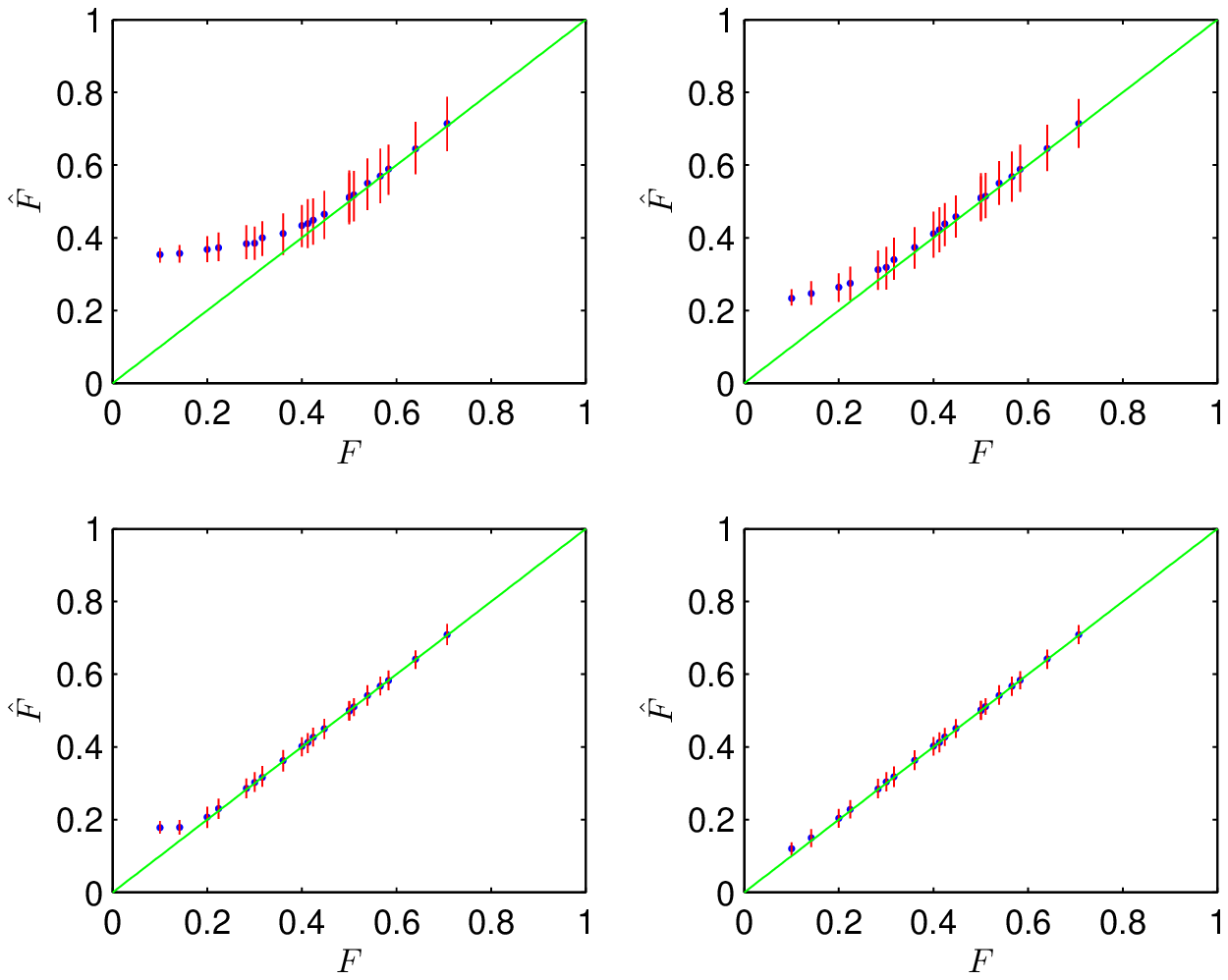}
\caption{Non-blind detection. Estimated flux of the polarized
sources
  $\hat{F}$ in Jy plotted against their real flux $F$.  The average
  and $68\%$ confidence intervals (vertical bar) of 500 simulations
  are plotted. Top left: the NPF has been used and the noise
  corresponds to the high noise zone. Top right: FF and high
  noise. Bottom left: NPF and low noise . Bottom right: FF and low
  noise. In all the plots the straight line $\hat{F}=F$ is drawn for
  comparison.} \label{fig:figpola4}
\end{figure*}

We carry out simulations with a polarized source placed in the image
centre and filter the images with the two different filters,
centering them in the pixels at a distance from the source less than
one FWHM. We simulate the same range of $(F_Q,F_U)$ pairs as in the
blind case (section~\ref{sec:blind}). Then, as in the blind case, we
calculate the maximum in $A$, using only the selected pixels. This
maximum is the estimated polarization of the source $\hat{A}$ and
the pixel where we find the maximum is the source position, we
convert this amplitude to the estimated flux in Jy, $\hat{F}$. We
have also performed a detection test, accepting as real sources only
those detected with a significance lower than $\alpha=0.05$. In
Table~\ref{tb:table2}, we write the power of the test, the average
of the relative polarization error, of its absolute value and of the
position error. Our results are obtained from 500 simulations for
each combination of pairs $(F_Q,F_U)$ with high and low noise.  The
detection power is higher for the FF than for the other filter. For
the high noise case, the improvement is very clear for $F \leq 0.50$
Jy; we also obtain higher powers for the non-blind case than for the
blind one. This could be expected, since we know the source position
in the former case.

\begin{figure*}
\centering
\includegraphics{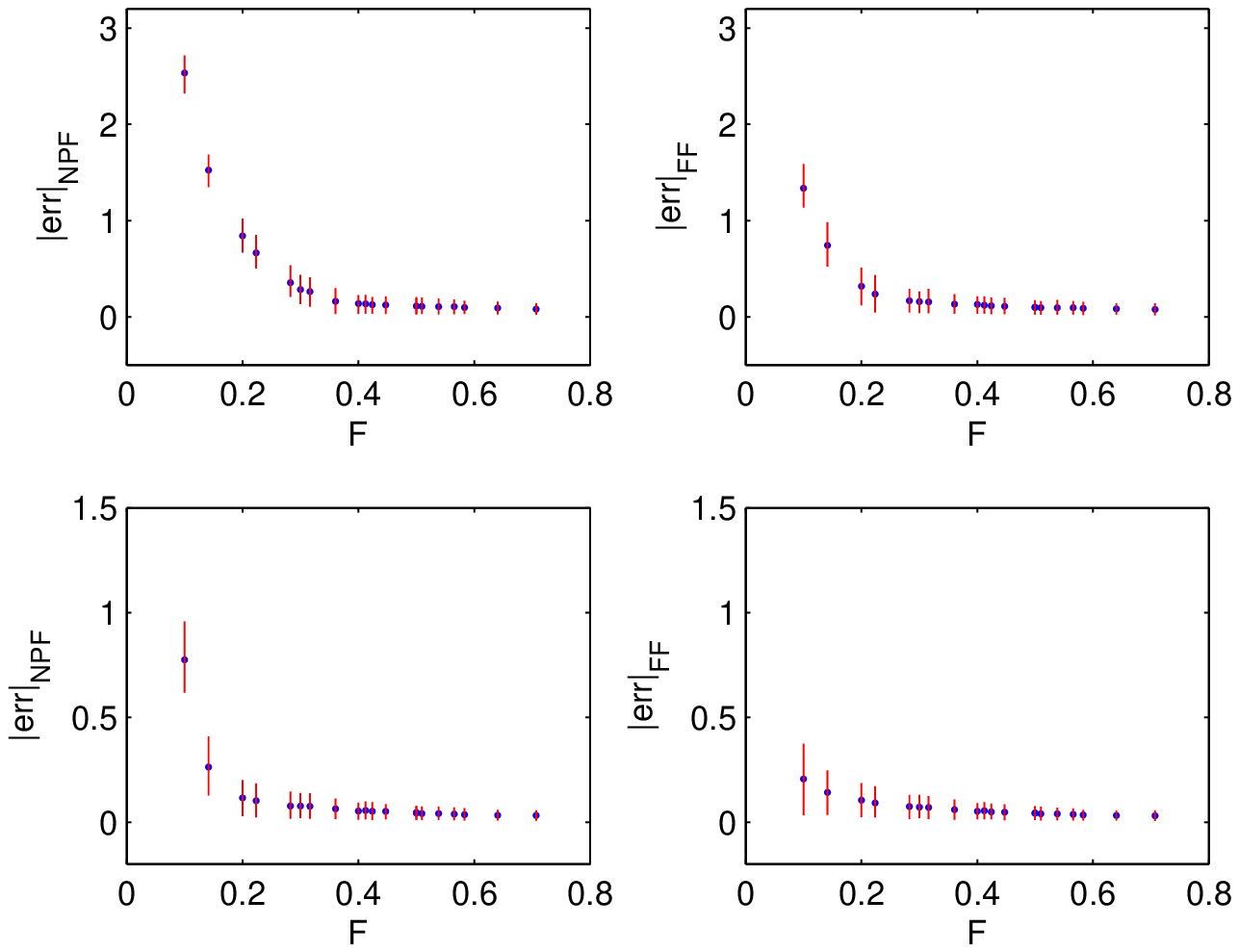}
\caption{Non-blind detection. Absolute value of the relative error,
  $|$err$|$, in the estimation of the flux of polarized sources
  plotted against the real flux $F$ in Jy. The average and $68\%$
  confidence intervals (vertical bar) of 500 simulations are
  plotted. Top left: The NPF is used with high noise. Top right: FF
  and high noise. Bottom left: NPF and low noise. Bottom right: FF and
  low noise.} \label{fig:figpola5}
\end{figure*}

 The position and polarization errors are also
lower for the FF, this can also be seen in Figs.~\ref{fig:figpola4}
and~\ref{fig:figpola5}, which show the same quantities as
Figs.~\ref{fig:figpola1} and~\ref{fig:figpola2} for non-blind
detection. For fluxes higher than $0.36$ Jy, the power with the FF
is $100\%$, the bias is $\leq 0.03$,the average of the absolute
value of the relative error is $\leq 0.13$ and the average position
error is $\leq 0.55$. The flux and position errors are lower in the
non-blind case than in the blind one, especially for low fluxes.

 The results for the low noise case are much better, the FF outperforms the NPF. From $0.28 $ Jy on
the results are quite similar. For fluxes higher than $0.18 $ Jy,
the power with the FF is $100\%$, the average bias is $\leq 0.02$,
the average of the absolute value of the relative error is $\leq
0.11$ and the average position error is $\leq 0.39$. Similarly to
what happened with the low noise case of Table~\ref{tb:table1}, we
have cut the Table at flux $F=0.40$ Jy and we have added new cases
in order to have a better sampling of the interesting signal to
noise regime. In this case, we have simulated the pairs
$(F_Q,F_U)=(0.05,0.10)$, $(0.00,0.12)$, $(0.00,0.13)$,
$(0.00,0.15)$, $(0.10,0.15)$ and $(0.00,0.25)$ Jy. The non-blind
case gives better results than the blind one, especially for low
fluxes.

\section{Conclusions} \label{sec:conclusions}

In this paper, we deal with the detection and estimation of the
modulus of a vector, a problem of great interest in general and in
particular in astrophysics when we consider the polarization of the
cosmic microwave background (CMB). The polarization intensity $P$ is
defined as $P\equiv (Q^2+U^2+V^2)^{\frac{1}{2}}$, where $Q$, $U$ and
$V$ are the Stokes parameters. We consider the case of images in
$Q$, $U$ and $V$ consisting of a compact source with a profile
$\tau(\vec{x})$ immersed in Gaussian uncorrelated noise. We intend
to detect the source and estimate its amplitude in $P$ by using two
different methods: a Neyman-Pearson filter (NPF) operating in $P$
and based on the maximisation of the corresponding log-likelihood
and a filtered fusion (FF) procedure, i.e. the application of the MF
on the images of $Q$, $U$ and $V$ and the combination of the
corresponding estimates by making the non-linear fusion $P\equiv
(Q_{MF}^2+U_{MF}^2+V_{MF}^2)^{\frac{1}{2}}$. We present the two
filters in section~\ref{sec:methodology} for two-dimensional, $V=0$,
and three-dimensional vectors, deriving the corresponding
expressions for the estimation of the polarized source amplitude.

Since we are interested in applying these different filters to the
CMB and this radiation is linearly polarized, we will only consider
the two-dimensional vector case in our simulations.

Our goal is to compare the performance of the filters when applied to
simulated images. Then, we carry out two-dimensional simulations with
the characteristics of the 70 GHz \emph{Planck} channel. The images
have $16\times 16$ pixels with a pixel size of $3.43'$. We simulate
the $Q$ and $U$ components consisting of a compact source filtered
with a Gaussian-shaped beam (FWHM of 14') plus CMB and non-stationary
detector noise. We consider two typical zones of the \emph{Planck}
survey: one with high noise and quite isotropic and another one with
low noise but proportionally more anisotropic. These zones are extreme
cases for the assumed scanning strategy we have chosen, and any other
zone of the sky is an intermediate case between these two.

We study two types of detection: blind detection, in which we do not
know the source position and non-blind, in which the position is
known; we place the source at the centre of the image. We take
values of the source fluxes in $Q$ and $U$, $F_Q$, and $F_U$,
ranging from 0.1 Jy to 0.5 Jy with a step of 0.1 Jy. Note that for
extragalactic objects both $Q$ and $U$ can take negative values, but
since the sign of both components is irrelevant for the calculation
of $P$ here we only simulate the positive case. We carry out 500
simulations for each pair $(F_Q,F_U)$ and for all the cases of blind
and non-blind detection with high and low noise.

We apply the filters to the simulated images, estimating the source
amplitude $A$ as the maximum value of $P$ in the filtered images and
the source position as the position of this maximum, then we convert
the source amplitude to the source flux in Jy . We fix a significance
$\alpha=0.05$ for the detection and calculate a detection power for
this significance, see section~\ref{sec:simulations} for more
details. We also calculate the relative error of the estimated flux,
its absolute value and the position error (in number of pixels), these
errors together with the detection power are written in
Tables~\ref{tb:table1} and~\ref{tb:table2} for the blind and non-blind
case. We also show the estimated fluxes and the absolute value of the
relative errors in
Figs.~\ref{fig:figpola1},~\ref{fig:figpola2},~\ref{fig:figpola4}
and~\ref{fig:figpola5} for the different cases.

The FF performs better than the NPF (as can be seen in the Tables and
Figures), especially for low fluxes and it is also much faster than
the NPF. However NPF is still interesting in a case where only the
modulus of a vector is known and not its components (for example, if
we had a map of $P$ polarization but not the $Q$ and $U$ maps
separately).  The powers are much higher and the errors much lower for
the low noise case than for the high noise one. The filters also
perform better in the non-blind case than in the blind one, especially
for low fluxes.

We can detect extragalactic point sources in polarization images (at
100\% power) with the FF in the high noise zone with fluxes $\geq
(0.42,0.36)$ Jy for (blind/non-blind) detection and in the low noise
zone with fluxes $\geq (0.22,0.18)$ Jy for (blind/non-blind)
detection. The bias and the position error are very low $\leq 0.03$
and $< 1$ pixel, respectively for all these fluxes.

\begin{table*}
  \centering
  \caption{BLIND DETECTION. First column: pairs of values of $F_Q$ and
    $F_U$ in Jy and the corresponding value of $F$ for the high-noise
    zone (top) and the low-noise zone (bottom). Second and third
    columns: detection power for the NPF and FF (percentage). Fourth
    and fifth columns: flux relative errors (average from 500
    simulations) for the two filters. Sixth and seventh columns:
    absolute value of the relative error (average from 500
    simulations). Eighth and ninth columns: position errors in numbers
    of pixels (average from 500 simulations).}
  \label{tb:table1}

  \begin{tabular}{ccccccccc}

      \textbf{High noise ($F_Q$,$F_U$;$F$)}& \textbf{pow$_{NPF}$}&
      \textbf{pow$_{FF}$} &\textbf{err$_{NPF}$}& \textbf{err$_{FF}$} &
      \textbf{$|$err$|$$_{NPF}$} &  \textbf{$|$err$|$$_{FF}$} &
      \textbf{pos$_{NPF}$} &  \textbf{pos$_{FF}$}
      \\

      \hline\hline

      \textbf{(0.00,0.10;0.10)} & 5 & 9 & 3.00 & 1.65 & 3.00 & 1.65 &
      7.84 & 4.31 \\ \textbf{(0.10,0.10;0.14)} & 8 & 17 & 1.87 & 0.97
      & 1.87 & 0.97 & 7.97 & 2.74 \\ \textbf{(0.00,0.20;0.20)} & 9 &
      43 & 1.06 & 0.45 & 1.06 & 0.45 & 4.57 & 1.47 \\
      \textbf{(0.10,0.20;0.22)} & 13 & 54 & 0.86 & 0.32 & 0.86 & 0.32
      & 4.40 & 1.13 \\ \textbf{(0.20,0.20;0.28)} & 22 & 81 & 0.48 &
      0.13 & 0.48 & 0.16 & 2.32 & 0.88 \\ \textbf{(0.00,0.30;0.30)} &
      29 & 91 & 0.43 & 0.12 & 0.43 & 0.16 & 1.73 & 0.73 \\
      \textbf{(0.10,0.30;0.32)} & 37 & 94 & 0.35 & 0.10 & 0.35 & 0.15
      & 2.23 & 0.63 \\ \textbf{(0.20,0.30;0.36)} & 52 & 99 & 0.22 &
      0.05 & 0.22 & 0.13 & 1.26 & 0.48 \\ \textbf{(0.00,0.40;0.40)} &
      64 & 99 & 0.14 & 0.02 & 0.15 & 0.13 & 1.11 & 0.46 \\
      \textbf{(0.10,0.40;0.41)} & 73 & 99 & 0.11 & 0.01 & 0.12 & 0.11
      & 1.18 & 0.40 \\ \textbf{(0.30,0.30;0.42)} & 78 & 100 & 0.10 &
      0.03 & 0.13 & 0.12 & 0.92 & 0.36 \\ \textbf{(0.20,0.40;0.45)} &
      87 & 100 & 0.07 & 0.03 & 0.11 & 0.11 & 0.97 & 0.28 \\
      \textbf{(0.00,0.50;0.50)} & 93 & 100 & 0.02 & 0.01 & 0.10 & 0.10
      & 0.59 & 0.28 \\ \textbf{(0.30,0.40;0.50)} & 97 & 100 & 0.03 &
      0.02 & 0.11 & 0.10 & 0.67 & 0.29 \\ \textbf{(0.10,0.50;0.51)} &
      95 & 100 & 0.02 & 0.01 & 0.11 & 0.10 & 0.58 & 0.27 \\
      \textbf{(0.20,0.50;0.54)} & 98 & 100 & 0.02 & 0.01 & 0.10 & 0.09
      & 0.51 & 0.17 \\ \textbf{(0.40,0.40;0.57)} & 99 & 100 & 0.01 &
      0.00 & 0.10 & 0.09 & 0.39 & 0.17 \\ \textbf{(0.30,0.50;0.58)} &
      100 & 100 & 0.01 & 0.01 & 0.10 & 0.09 & 0.41 & 0.19 \\
      \textbf{(0.40,0.50;0.64)} & 100 & 100 & 0.01 & 0.01 & 0.09 &
      0.08 & 0.27 & 0.10 \\ \textbf{(0.50,0.50;0.71)} & 100 & 100 &
      0.00 & 0.00 & 0.08 & 0.07 & 0.20 & 0.10 \\

      \hline\hline

      \textbf{Low noise ($F_Q$,$F_U$;$F$)}& \textbf{pow$_{NPF}$}&
      \textbf{pow$_{FF}$} &\textbf{err$_{NPF}$}& \textbf{err$_{FF}$} &
      \textbf{$|$err$|$ $_{NPF}$} &  \textbf{$|$err$|$$_{FF}$} &
      \textbf{pos$_{NPF}$} &  \textbf{pos$_{FF}$}
      \\

      \hline\hline

      \textbf{(0.00,0.10;0.10)} & 5 & 39 & 1.01 & 0.37 & 1.01 & 0.37 &
      4.53 & 1.57 \\ \textbf{(0.05,0.10;0.11)} & 9 & 53 & 0.81 & 0.23
      & 0.81 & 0.23 & 5.04 & 1.16 \\ \textbf{(0.10,0.10;0.14)} & 17 &
      91 & 0.45 & 0.08 & 0.45 & 0.13 & 2.54 & 0.55
      \\ \textbf{(0.00,0.15;0.15)} & 19 & 93 & 0.39 & 0.07 & 0.39 &
      0.13 & 1.94 & 0.48 \\ \textbf{(0.05,0.15;0.16)} & 27 & 97 & 0.32
      & 0.05 & 0.32 & 0.13 & 2.13 & 0.45 \\ \textbf{(0.10,0.15;0.18)}
      & 55 & 99 & 0.16 & 0.03 & 0.16 & 0.12 & 1.30 & 0.34
      \\ \textbf{(0.00,0.20;0.20)} & 79 & 99 & 0.08 & 0.03 & 0.10 &
      0.10 & 0.71 & 0.24 \\ \textbf{(0.10,0.20;0.22)} & 92 & 100 &
      0.04 & 0.01 & 0.09 & 0.09 & 0.45 & 0.21
      \\ \textbf{(0.00,0.25;0.25)} & 99 & 100 & 0.00 & 0.00 & 0.08 &
      0.08 & 0.43 & 0.16 \\ \textbf{(0.20,0.20;0.28)} & 100 & 100 &
      0.01 & 0.01 & 0.08 & 0.07 & 0.18 & 0.09
      \\ \textbf{(0.00,0.30;0.30)} & 100 & 100 & 0.00 & 0.00 & 0.08 &
      0.07 & 0.20 & 0.10 \\ \textbf{(0.10,0.30;0.32)} & 100 & 100 &
      0.01 & 0.01 & 0.07 & 0.06 & 0.15 & 0.08
      \\ \textbf{(0.20,0.30;0.36)} & 100 & 100 & 0.00 & 0.00 & 0.06 &
      0.06 & 0.10 & 0.05 \\ \textbf{(0.00,0.40;0.40)} & 100 & 100 &
      0.00 & 0.00 & 0.06 & 0.05 & 0.05 & 0.03
      \\

      \hline\hline
  \end{tabular}
\end{table*}

\begin{table*}
  \centering
  \caption{NON-BLIND DETECTION. First column: pairs of values of $F_Q$
    and $F_U$ in Jy and the corresponding value of $F$ for the
    high-noise zone (top) and the low-noise zone (bottom). Second and
    third columns: detection power for the NPF and FF
    (percentage). Fourth and fifth columns: flux relative errors
    (average from 500 simulations) for the two filters. Sixth and
    seventh columns: absolute value of the relative error (average
    from 500 simulations). Eighth and ninth columns: position errors
    in numbers of pixels (average from 500 simulations).}
  \label{tb:table2}

  \begin{tabular}{ccccccccc}

    \textbf{High noise ($F_Q$,$F_U$;$F$)}& \textbf{pow$_{NPF}$}&
    \textbf{pow$_{FF}$} &\textbf{err$_{NPF}$}& \textbf{err$_{FF}$} &
    \textbf{$|$err$|$$_{NPF}$} & \textbf{$|$err$|$$_{FF}$} &
    \textbf{pos$_{NPF}$} & \textbf{pos$_{FF}$} \\

    \hline\hline

    \textbf{(0.00,0.10;0.10)} & 9 & 19 & 2.53 & 1.33 & 2.53 & 1.33 &
    1.90 & 1.52 \\ \textbf{(0.10,0.10;0.14)} & 13 & 35 & 1.52 & 0.74 &
    1.52 & 0.74 & 1.70 & 1.37 \\ \textbf{(0.00,0.20;0.20)} & 22 & 68 &
    0.84 & 0.32 & 0.84 & 0.32 & 1.53 & 1.09 \\
    \textbf{(0.10,0.20;0.22)} & 32 & 78 & 0.66 & 0.23 & 0.66 & 0.24 &
    1.44 & 0.90 \\ \textbf{(0.20,0.20;0.28)} & 52 & 95 & 0.35 & 0.10 &
    0.35 & 0.17 & 1.26 & 0.81 \\ \textbf{(0.00,0.30;0.30)} & 53 & 99 &
    0.28 & 0.06 & 0.28 & 0.16 & 1.18 & 0.67 \\
    \textbf{(0.10,0.30;0.32)} & 63 & 98 & 0.26 & 0.07 & 0.26 & 0.15 &
    1.13 & 0.66 \\ \textbf{(0.20,0.30;0.36)} & 77 & 100 & 0.14 & 0.03
    & 0.16 & 0.13 & 0.92 & 0.55 \\ \textbf{(0.00,0.40;0.40)} & 90 &
    100 & 0.08 & 0.03 & 0.14 & 0.13 & 0.79 & 0.38 \\
    \textbf{(0.10,0.40;0.41)} & 92 & 100 & 0.06 & 0.02 & 0.13 & 0.12 &
    0.77 & 0.39 \\ \textbf{(0.30,0.30;0.42)} & 94 & 100 & 0.06 & 0.03
    & 0.13 & 0.12 & 0.77 & 0.39 \\ \textbf{(0.20,0.40;0.45)} & 97 &
    100 & 0.04 & 0.02 & 0.12 & 0.11 & 0.72 & 0.40 \\
    \textbf{(0.00,0.50;0.50)} & 99 & 100 & 0.02 & 0.02 & 0.11 & 0.10 &
    0.58 & 0.25 \\ \textbf{(0.30,0.40;0.50)} & 100 & 100 & 0.02 & 0.02
    & 0.11 & 0.10 & 0.59 & 0.29 \\ \textbf{(0.10,0.50;0.51)} & 99 &
    100 & 0.02 & 0.01 & 0.11 & 0.09 & 0.59 & 0.30 \\
    \textbf{(0.20,0.50;0.54)} & 100 & 100 & 0.02 & 0.02 & 0.11 & 0.09
    & 0.49 & 0.22 \\ \textbf{(0.40,0.40;0.57)} & 100 & 100 & 0.01 &
    0.00 & 0.10 & 0.09 & 0.40 & 0.18 \\ \textbf{(0.30,0.50;0.58)} &
    100 & 100 & 0.01 & 0.01 & 0.10 & 0.09 & 0.43 & 0.16 \\
    \textbf{(0.40,0.50;0.64)} & 100 & 100 & 0.01 & 0.01 & 0.09 & 0.08
    & 0.30 & 0.14 \\ \textbf{(0.50,0.50;0.71)} & 100 & 100 & 0.01 &
    0.01 & 0.08 & 0.08 & 0.23 & 0.11 \\

    \hline\hline

    \textbf{Low noise ($F_Q$,$F_U$;$F$)}& \textbf{pow$_{NPF}$}&
    \textbf{pow$_{FF}$} &\textbf{err$_{NPF}$}& \textbf{err$_{FF}$} &
    \textbf{$|$err$|$$_{NPF}$} & \textbf{$|$err$|$$_{FF}$} &
    \textbf{pos$_{NPF}$} & \textbf{pos$_{FF}$} \\

    \hline\hline

    \textbf{(0.00,0.10;0.10)} & 12 & 69 & 0.77 & 0.20 & 0.77 & 0.21 &
    1.44 & 0.78 \\ \textbf{(0.05,0.10;0.11)} & 22 & 88 & 0.58 & 0.15 &
    0.58 & 0.18 & 1.52 & 0.72 \\ \textbf{(0.00,0.12;0.12)} & 24 & 93 &
    0.47 & 0.10 & 0.47 & 0.15 & 1.18 & 0.65
    \\ \textbf{(0.00,0.13;0.13)} & 34 & 95 & 0.38 & 0.09 & 0.38 & 0.16
    & 1.14 & 0.57 \\ \textbf{(0.10,0.10;0.14)} & 48 & 98 & 0.26 & 0.06
    & 0.26 & 0.14 & 0.92 & 0.51 \\ \textbf{(0.00,0.15;0.15)} & 57 & 99
    & 0.21 & 0.05 & 0.21 & 0.14 & 0.82 & 0.52
    \\ \textbf{(0.10,0.15;0.18)} & 87 & 100 & 0.07 & 0.01 & 0.11 &
    0.11 & 0.61 & 0.39 \\ \textbf{(0.00,0.20;0.20)} & 94 & 100 & 0.03
    & 0.02 & 0.12 & 0.11 & 0.56 & 0.28 \\ \textbf{(0.10,0.20;0.22)} &
    99 & 100 & 0.03 & 0.02 & 0.10 & 0.09 & 0.39 & 0.23
    \\ \textbf{(0.00,0.25;0.25)} & 100 & 100 & 0.02 & 0.01 & 0.09 &
    0.08 & 0.32 & 0.15 \\ \textbf{(0.20,0.20;0.28)} & 100 & 100 & 0.01
    & 0.01 & 0.08 & 0.07 & 0.22 & 0.11 \\ \textbf{(0.00,0.30;0.30)} &
    100 & 100 & 0.01 & 0.01 & 0.08 & 0.07 & 0.21 & 0.09
    \\ \textbf{(0.10,0.30;0.32)} & 100 & 100 & 0.00 & 0.00 & 0.08 &
    0.07 & 0.16 & 0.08 \\ \textbf{(0.20,0.30;0.36)} & 100 & 100 & 0.00
    & 0.01 & 0.06 & 0.06 & 0.09 & 0.04 \\ \textbf{(0.00,0.40;0.40)} &
    100 & 100 & 0.00 & 0.01 & 0.05 & 0.05 & 0.08 & 0.04 \\

    \hline\hline

  \end{tabular}

\end{table*}

\section*{Acknowledgements}

The authors acknowledge partial financial support from the Spanish
Ministry of Education (MEC) under project ESP2004-07067-C03-01 and
from the joint CNR-CSIC research project 2006-IT-0037. JLS
acknowledges partial financial support by the Spanish MEC and thanks
the CNR ISTI in Pisa for their hospitality during his sabbatical
leave. FA also wishes to thank the CNR ISTI in Pisa for their
hospitality during two short stays when part of this work was done.
MLC acknowledges the Spanish MEC for a postdoctoral fellowship. Some
of the results in this paper have been derived using the
HEALPix~\citep{healpix} package.

\bibliographystyle{mn2e}
\bibliography{mibib}

\label{lastpage}

\end{document}